\pdfoutput=1
\documentclass[aps,prl,twocolumn,nofootinbib]{revtex4-2}
\usepackage{pdfpages}
\makeatletter\AtBeginDocument{\let\LS@rot\@undefined}\makeatother

\usepackage{amsmath}
\usepackage{amsfonts}
\usepackage{amssymb}
\usepackage{graphicx}
\usepackage{bm}
\usepackage{color}
\usepackage{ulem}
\usepackage{commath}
\usepackage[T1]{fontenc}
\usepackage{lmodern}
\usepackage{xcolor}
\usepackage{hyperref}
\usepackage{orcidlink}
\usepackage{comment}
\allowdisplaybreaks

\def\TEOBResumS{\texttt{TEOBResumS}}
\def\Dali{\texttt{TEOBResumS-Dal\'i}}
\def\LMR{\texttt{TEOB-LMR}}

\begin{document}

\title{An uncalibrated effective-one-body waveform model \\
for intermediate-mass-ratio black hole binaries}

\author{Angelica \surname{Albertini}\orcidlink{0000-0002-9556-1323}}
\author{Vojt\v{e}ch \surname{Witzany}\orcidlink{0000-0002-9209-5355}}

\affiliation{Institute of Theoretical Physics,
Faculty of Mathematics and Physics, Charles University,
V Holešovičkách 2,
180 00, Praha 8, Czech Republic}

\begin{abstract}
Intermediate-mass-ratio binaries, with mass ratios $q=m_1/m_2\sim10^2$--$10^4$, are among the most challenging sources for waveform modeling. We present an inspiral-merger-ringdown effective-one-body model for nonspinning, quasicircular binaries with $q>10$, with dynamics built only from analytical post-Newtonian results. It reaches $(2,2)$ unfaithfulness of $10^{-4}$--$10^{-5}$ against numerical-relativity simulations with $q>10$ and subradian dephasing against second-order self-force waveforms over four-year inspirals at $q=10^4$--$10^5$. An optimized post-adiabatic
implementation enables waveform generation up to two orders of magnitude
faster than integrating the full equations of motion.
\end{abstract}

\maketitle

\paragraph{Introduction.}
Gravitational-wave astronomy will soon span an enormous range of masses and mass
ratios. Advanced LIGO~\cite{LIGOScientific:2014pky}, Virgo~\cite{VIRGO:2014yos}
and KAGRA~\cite{KAGRA:2020tym} observe stellar-mass binaries, mostly of comparable
mass. Next-generation detectors, the Einstein Telescope~\cite{Punturo:2010zz,ET:2019dnz,Branchesi:2023mws}
and Cosmic Explorer~\cite{Reitze:2019iox,Evans:2021gyd} on the ground and
LISA~\cite{LISA:2017pwj,LISA:2024hlh}, Taiji~\cite{Hu:2017mde} and TianQin~\cite{TianQin:2015yph} in space,
will open a window on new sources: massive-black-hole binaries, extreme-mass-ratio inspirals (EMRIs),
with mass ratios $q\equiv m_1/m_2\gtrsim10^4$, and, in between, intermediate-mass-ratio binaries
(IMRBs), with $q\sim10^2$--$10^4$, which involve intermediate-mass black holes
(IMBHs, $10^2$--$10^5\,M_\odot$) and are too costly for numerical relativity (NR).

Such binaries come in two kinds~\cite{LISA:2024hlh}. In light IMRBs, a stellar-mass compact object
spirals into an IMBH.
The IMBH grows in a dense star cluster, through stellar collisions or repeated
black-hole mergers, and mass segregation then drives compact objects onto
it~\cite{Miller:2001ez,Amaro-Seoane:2012lgq}. LISA can detect light IMRBs with
IMBHs of $10^3$--$10^4\,M_\odot$~\cite{Meus:2026owb}; lighter ones radiate in the decihertz and
ground-based bands~\cite{Amaro-Seoane:2018gbb,Sedda:2019uro,Arca-Sedda:2020lso}.
In heavy IMRBs, the IMBH is the lighter body, spiraling into a massive black hole
in a galactic center after an infalling star cluster delivers it
there~\cite{LISA:2024hlh}. Both kinds can coalesce in band, so their templates must include merger and ringdown.

Comparable-mass waveform models~\cite{Varma:2019csw,Estelles:2021gvs,Thompson:2023ase,Nagar:2024oyk,Colleoni:2024knd,Hamilton:2025xru,Gamboa:2026jht} degrade with growing mass ratio~\cite{KAGRA:2021vkt,Mahapatra:2026wsp}. Self-force models, expanding in $1/q$ around the test-mass
limit~\cite{Barack:2018yvs,Pound:2019lzj}, are reliable for EMRIs but lack the higher-order mass-ratio information needed in between, and are mostly inspiral-only, despite recent progress on plunge and merger~\cite{Kuchler:2024esj,Kuchler:2025hwx}. What is needed is a model that interpolates between the two ends, where the information exists.

Several recent models reach into this regime, each limited by its input
information. The hybrid \texttt{WaSABI-C}~\cite{Honet:2025gge,Honet:2025lmk} assembles
first-order self-force (1SF) and post-Newtonian (PN) results without NR tuning, but covers the inspiral
only. The surrogates \texttt{BHPTNRSur1dq1e4}~\cite{Islam:2022laz} and
\texttt{BHPTNRSur2dq1e3}~\cite{Rink:2024swg}, trained on perturbation-theory waveforms
calibrated to NR, and \texttt{WaSABINRHybSur2dq15}~\cite{Iglesias:2026ypj}, trained on
hybrids of \texttt{WaSABI-C} and NR up to $q=20$, are limited in the covered frequency range. %
\texttt{SEOBNR-GSF}~\cite{Leather:2025nhu}, with a self-force-informed Hamiltonian, is benchmarked against NR only, so its performance for IMRBs is unknown. %
A calibration to NR or numerical self-force data constrains a model only where those data exist, which is precisely not the intermediate regime. We show that \textit{currently known analytical information only}, properly resummed, already gives a satisfactory model there.

The effective-one-body (EOB) formalism is built to interpolate between the two
ends~\cite{Buonanno:1998gg,Buonanno:2000ef,Damour:2008yg}: it maps the two-body
dynamics onto a test body in a deformed background by resumming PN results, and has an exact test-mass limit by construction.
We work with the EOB model \TEOBResumS{}~\cite{Damour:2014sva,Nagar:2018zoe}, in its most recent version \Dali{}, built for generic orbits~\cite{Nagar:2024oyk,Albanesi:2025txj} and calibrated to NR at comparable masses.
Refs.~\cite{Albertini:2022dmc,Albertini:2023aol,Albertini:2024rrs} presented a large-mass-ratio version of \TEOBResumS{}, aimed at EMRIs, but its self-force-informed potentials are singular at the light ring in the Damour-Jaranowski-Sch\"afer (DJS) gauge~\cite{Damour:2000we,Nagar:2022fep,Akcay:2012ea},\footnote{For this reason Refs.~\cite{Antonelli:2019fmq,Leather:2025nhu} incorporate self-force
information into EOB Hamiltonians in a different gauge.} restricting it to the inspiral, and its long ODE integrations made it slow.

Here we target the intermediate regime and remove both limitations. Analytical information assembled into the \Dali{} potentials, with an improved radiation reaction, makes the model faithful at both ends without calibration: to NR at $10<q\le20$ and to self-force waveforms at $q=10^4$--$10^5$. A post-adiabatic inspiral extended to where it converges, and a fast waveform path, make the cost of full frequency-domain waveforms nearly independent of the mass ratio.

\paragraph{Quick EOB reminder.}
We use $G=c=1$, total mass $M=m_1+m_2$, reduced mass $\mu=m_1m_2/M$, symmetric mass ratio $\nu =q/(1+q)^2$, and
dimensionless phase-space variables: the EOB radial separation $r=R/M$, with
$u\equiv1/r$, the momenta $p_{r_*}=P_{R_*}/\mu$ and $p_\varphi=P_\varphi/(\mu M)$,
and time $t=T/M$. The conservative
dynamics follows from
$\hat{H}_{\rm EOB}\equiv H_{\rm EOB}/\mu = \left(\sqrt{1+2\nu(\hat H_{\rm
eff}-1)}\right)/\nu$, with the nonspinning effective Hamiltonian
\begin{equation}
\label{eq:Heff-nonspin}
\hat H_{\rm eff}\equiv \frac{H_{\rm eff}}{\mu}=\sqrt{p_{r_*}^2 + A(u)\left[1+p_\varphi^2 u^2
+ Q^*(u,p_{r_*})\right]},
\end{equation}
where $p_\varphi$ is the angular momentum and $p_{r_*}$ the momentum conjugate
to the tortoise radial coordinate, $p_{r_*} = \left( A/B \right)^{1/2} p_r $.
The three radial potentials, defined in the DJS
gauge~\cite{Damour:2000we}, are the main potential $A(u;\nu)\to1-2u$
(test-mass limit), $B(u;\nu)$, usually recast as
$\bar D(u;\nu)\equiv1/D\equiv1/(AB)$, and $Q^*(u,p_{r_*};\nu)$. $A$ and $B$ are the $tt$ and $rr$ coefficients of the effective metric; $Q$, from 3PN
on~\cite{Damour:2000we}, deforms the mass-shell condition, and $Q^*$ is $Q$ written in terms of $p_{r_*}$.

The equations of motion for quasicircular binaries read
\begin{align}
\label{eq:eom}
\dot r &= \left(\frac{A}{B}\right)^{1/2}
          \frac{\partial \hat H_{\rm EOB}}{\partial p_{r_*}}, &
\dot p_{r_*} &= -\left(\frac{A}{B}\right)^{1/2}
          \frac{\partial \hat H_{\rm EOB}}{\partial r}, \nonumber\\
\dot \varphi &= \frac{\partial \hat H_{\rm EOB}}{\partial p_\varphi}
          \equiv \Omega, &
\dot p_\varphi &= \hat{\mathcal F}_\varphi .
\end{align}

The flux is the resummed multipolar expression of
Refs.~\cite{Damour:2008gu,Nagar:2011aa},
\begin{align}
\label{eq:flux}
\hat{\mathcal F}_\varphi &= -\frac{32}{5}\,\nu\,r_\omega^4\,\Omega^5
   \left[\hat f(x) + \hat f_{\rm H}(x)\right],\\
\hat f(x) &= \frac{1}{{\cal F}^{\rm N}_{22}}\sum_{\ell m} {\cal F}^{\rm N}_{\ell m}
   \big|\hat h_{\ell m}\big|^2 ,
\end{align}
where $\hat f_{\rm H}$ is the horizon flux, ${\cal F}^{\rm N}_{\ell m}$ are the Newtonian multipolar fluxes, $r_\omega$ is the
radius entering the generalized Kepler law $1 = \Omega^2r_\omega^3$~\cite{Damour:2006tr,Damour:2007xr}, $x=(r_\omega\Omega)^2$, and each
$\hat h_{\ell m}$ is factorized into an effective source ($\hat H_{\rm eff}$ or the Newton-normalized angular momentum $\hat j=p_\varphi/(r_\omega^2\Omega)$ for $\ell+m$ even or odd), a tail factor, a residual phase and a
residual amplitude correction $f_{\ell m}=\rho_{\ell m}^\ell$~\cite{Damour:2008gu}.

\paragraph{Waveform.}
The strain is decomposed on spin-weighted spherical harmonics,
$h_+-ih_\times=D_L^{-1}\sum_{\ell m}h_{\ell m}\,{}_{-2}Y_{\ell m}(\iota,\varphi_0)$,
with $D_L$ the luminosity distance and $(\iota,\varphi_0)$ the observer's
angles. Each multipole is written as $h_{\ell m}=A_{\ell m}e^{-i\phi_{\ell m}}$,
with frequency $\omega_{\ell m}\equiv\dot\phi_{\ell m}$.

\paragraph{The model.}
We refer to the model presented here as \texttt{TEOBResumS-Dal\'i-LMR}, shortened in the following to \LMR{}, for large mass ratio: it targets intermediate mass ratios but stays accurate for all $q>10$, into the extreme-mass-ratio regime. It is a branch of \Dali{}, from which it differs mainly in the radial potentials, the radiation reaction, the post-adiabatic step and handover, and the ODE tolerances; most of the rest is the baseline \Dali{} one (details in Table~S1 of the Supplemental Material).
The next-to-quasicircular corrections to the waveform are computed by matching the
amplitude and frequency at merger to NR-informed fits~\cite{Damour:2009kr,Riemenschneider:2021ppj,Nagar:2024oyk},
and the merger--ringdown is the NR-informed description of Refs.~\cite{Damour:2014yha,DelPozzo:2016kmd,Nagar:2019wds,Nagar:2020pcj,Pompili:2023tna,Nagar:2023zxh}.
The merger fits include the test-mass limit~\cite{Nagar:2019wds}, and the same merger--ringdown agrees with RIT
simulations up to $q=128$~\cite{Nagar:2022icd}.
These are the only NR-informed ingredients: no parameter of the dynamics is calibrated, and the remaining fits are test-mass fits of three subdominant horizon-flux multipoles.

\paragraph{Trial and error.}
Pad\'e resummations of the full high-order 1SF series~\cite{Bini:2014nfa,Bini:2015bla} develop poles at finite $\nu$. Keeping the \Dali{} potentials and calibrating $A$ on the second-order self-force model 1PAT1~\cite{Wardell:2021fyy} at $q=10^5$ and on NR at $q\le20$ was not successful either. The fit to 1PAT1 revealed a dephasing scaling as $1/\nu$, i.e.\ a test-mass (0th post-adiabatic, 0PA) flux error, which the modes with $\ell=11$--$13$ reduced.
We attribute the remaining, 1PA, dephasing to the analytically unknown 5PN $O(\nu)$ flux term. A fit of this term to 1PAT1 removes it, but must be damped at comparable masses with a guessed function: we abandoned this inelegant, overfitted model, and calibration altogether.

\paragraph{The radial potentials.}
We note two facts: (i) truncated at 5PN, $A$ misses the $O(\nu)$
binding energy given by the 1SF first law~\cite{LeTiec:2011ab} by $25\%$ at $r=6.5$;
(ii) its half-integer terms, known only at
$O(\nu)$, come in nearly canceling pairs: at $r=6$ the 5.5PN and 6.5PN terms
are $+2.7\%$ and $-2.4\%$ of the $O(\nu)$ contribution to $A$.  Keeping the
6PN term and the 5.5--6.5PN pair brings the binding energy within $0.7\%$ of the 1SF one. We therefore take
\begin{equation}
\label{eq:A_LMR}
A(u;\nu) = P^1_6\big[A_{6\rm PN}\big](u)
  + \nu\left(a_{6.5}\,u^{6.5} + a_{7.5}\,u^{7.5}\right),
\end{equation}
where $A_{6\rm PN}$ is the 6PN Taylor series with every known $\nu$-dependence,
including the 5PN $\nu^2$ term~\cite{Porto:2026fsd,Bini:2020wpo}, and the
logarithms are held fixed inside the Pad\'e. %
The half-integer pair stays outside:
resumming it breaks the cancellation and creates poles. %
The resulting $A$
has a single horizon and a single light ring at every $\nu$, and its $O(\nu)$ contribution lies
within $0.5\%$ of the exact 1SF function down to $r=5$ (End Matter, Fig.~\ref{fig:a1sf_reldiff}).
At comparable masses the
half-integer terms of Eq.~\eqref{eq:A_LMR}, linear in $\nu$, move the light ring at $\nu=1/4$ to $u\simeq0.36$, against $u\simeq0.55$ in \Dali{} (End Matter, Fig.~\ref{fig:u2A_LR}); we therefore restrict the model
to $q>10$, where its unfaithfulness against NR is $\sim10^{-4}$ or below.

The other potentials are:
\begin{itemize}
\item $\bar D$: the comparable-mass resummation of $D_{5\rm PN}$ from Ref.~\cite{Nagar:2024oyk},
completed with the analytical $5$PN
$\nu^2$ coefficient~\cite{Porto:2026fsd};
\item $Q^*$: the full $5$PN information, local plus nonlocal
($f$-route~\cite{Bini:2020wpo}), in every coefficient,
retaining the $p_{r_*}^4$ and $p_{r_*}^6$ sectors and dropping the $p_{r_*}^8$
one, whose nonlocal completion delays the plunge unphysically (full expression in the Supplemental Material).
\end{itemize}

\paragraph{Radiation reaction.}
The previous version~\cite{Albertini:2022dmc,Albertini:2023aol,Albertini:2024rrs} already used $22$PN test-mass $\rho_{\ell m}$ through $\ell=10$ and a multipolar ($\ell=2,3,4$) horizon flux. We add:
\begin{itemize}
    \item All the known $\nu$-dependent terms, including the 4PN $\rho_{22}$ coefficient from Ref.~\cite{Blanchet:2023sbv} (End Matter);
    \item Modes up to $\ell = 13$ with $m \ge \ell-2$, in the flux only;
    \item A horizon $\rho_{22}^{\rm H}$ resummed as $P^{30}_2$, and a test-mass fit for $\rho_{43}^{\rm H}$.
\end{itemize}

\paragraph{Speed.}
\label{sec:performance}
The ODE inspiral grows as $1/\nu$, which made the previous version~\cite{Albertini:2022dmc,Albertini:2023aol,Albertini:2024rrs} too slow for data analysis; two changes make the cost nearly independent of $q$.

First, the post-adiabatic iteration method~\cite{Nagar:2018gnk,Riemenschneider:2021ppj}
solves the quasicircular inspiral on a radial grid at $\nu$-independent cost, but the previous version~\cite{Albertini:2024rrs} stops it at $r=14$ and integrates the rest as ODEs. We
push the post-adiabatic inspiral down to the transition to
plunge~\cite{Ori:2000zn,Buonanno:2000ef,Compere:2021zfj,Kuchler:2024esj,RibesMetidieri:2026tat}
and switch to ODEs at $r_{\rm switch}=r_{\rm LSO}+c_{\rm sw}\,\Delta r_{\rm P}$, where
$r_{\rm LSO}$ is the last stable orbit (LSO) of the EOB dynamics at finite $\nu$,
$\Delta r_{\rm P}$ the width of the transition region, and $c_{\rm sw}$ a handover coefficient, fixed below. The width follows from the
right-hand sides of Eqs.~\eqref{eq:eom} at the LSO (End Matter),
\begin{align}
    \label{eq:DeltarP}
    \Delta r_{\mathrm{P}} &=
      \left( -\frac{1}{12}\,
        \frac{\partial \dot r}{\partial p_{r_*}}\,
        \frac{\partial^2 \dot p_{r_*}}{\partial r^2}
      \right)^{\!-3/5}
      \left( -\frac{\partial \dot r}{\partial p_{r_*}}\,
        \frac{\partial \dot p_{r_*}}{\partial p_\varphi}\, \hat{\mathcal F}_\varphi
      \right)^{\!2/5} \nonumber \\ &\propto \nu^{2/5},
\end{align}
with $\hat{\mathcal F}_\varphi\propto\nu$ the angular-momentum flux. With $c_{\rm sw}=\sqrt{4/3}$, the handover sits eight transition
timescales before the LSO crossing, and only ${\sim}600$ ODE steps remain at any
$\nu$. Against full ODE evolutions, the unfaithfulness stays below
$10^{-7}$, at speedups from $4\times$ ($q=20$) to ${\sim}300\times$
($q=10^5$) (End Matter, Fig.~\ref{fig:handover}). A semi-analytical transition-to-plunge expansion in $\nu^{1/5}$, following the theory of Refs.~\cite{Ori:2000zn,Buonanno:2000ef,Compere:2021zfj,Kuchler:2024esj,RibesMetidieri:2026tat}, instead diverged already for $\nu\gtrsim10^{-5}$ and was slower than the ODEs. All ODEs are integrated with
absolute and relative tolerances $10^{-14}$ and $10^{-12}$, ten times tighter
than in \Dali{}: this removes a spurious residual eccentricity at $q\sim10^5$
and halves the run time there.

Second, the inspiral multipoles are computed on coarse radial nodes and
interpolated (phase error $\lesssim4\times10^{-4}$~rad), by default in both domains.
In the frequency domain, a fast path also replaces the analytic prolongation of the
stationary-phase approximation above merger~\cite{Gamba:2020ljo} by a fast Fourier transform (FFT)
of the merger--ringdown. It costs $\sim60$~ms plus
$\sim0.1\,\mu$s per frequency bin at any $q$, with unfaithfulness against the
FFT of the full ODE evolution $\lesssim3\times10^{-4}$ at $q=20$, falling
below $10^{-6}$ for $q\ge10^4$. Up to $q\sim10^4$, a time-domain run with this handover
followed by an FFT costs the same and is more faithful ($\lesssim10^{-7}$), so it
is preferable there; the fast path pays off for the longest signals, ${\sim}5\times$ faster at $q=10^5$ (Supplemental Material).

\paragraph{Results.}
\label{sec:results}
We quantify agreement with the $(2,2)$
unfaithfulness $\bar{\mathcal F}=1-\max_{t_0,\phi_0}\langle h_1,h_2\rangle/(\|h_1\|\|h_2\|)$,
where $\langle\cdot,\cdot\rangle$ is the noise-weighted inner product, maximized over relative time and phase shifts; we compute it with
\texttt{PyART}\footnote{\url{https://github.com/RoxGamba/PyART}}. We use a private alignment code to evaluate the dephasing at merger (or at the final available point of the inspiral).

\begin{figure*}[t]
\includegraphics[width=2\columnwidth]{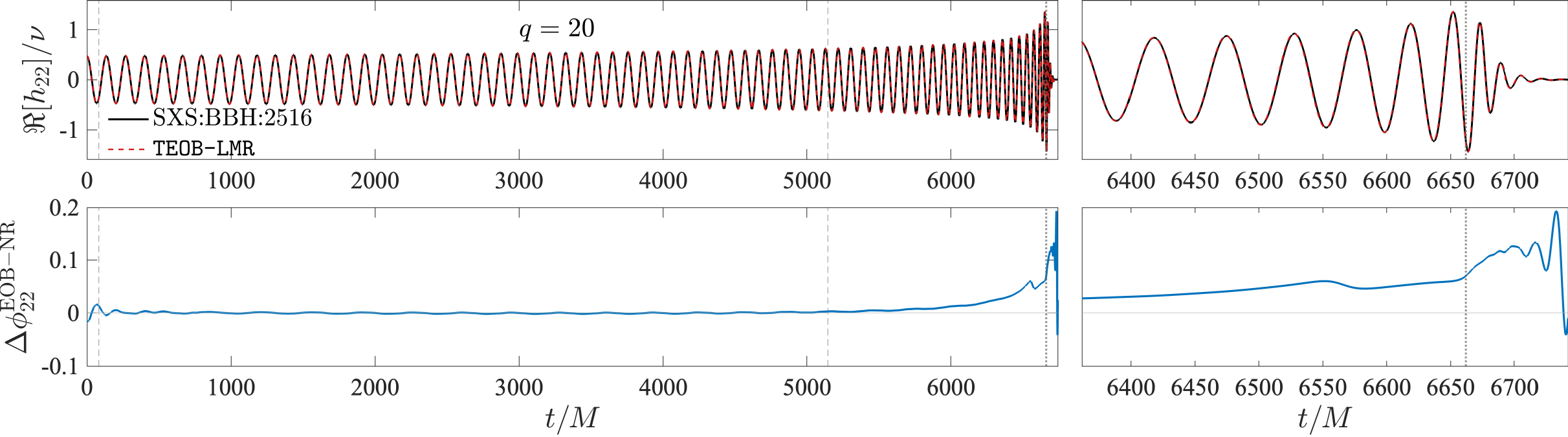}
\caption{\label{fig:dephasing_sxs} $(2,2)$ phase difference between \LMR{} and the
$q=20$ nonspinning SXS simulation SXS:BBH:2516, aligned over
$\omega_{22}\in[0.047,0.0705]$; residual alignment systematic $7\times10^{-4}$~rad.
\textit{Top:} real part of the $(2,2)$ strain, scaled by $\nu$; \textit{bottom:}
phase difference. \textit{Left:} the whole simulation; \textit{right:} zoom on
merger and ringdown. Dashed lines bound the alignment window; the dotted line
marks the merger.}
\end{figure*}

Table~\ref{tab:sxs_highq} lists the unfaithfulness against the nonspinning SXS
simulations~\cite{Scheel:2025jct} with $q\simeq10$--$20$. For $q>10$ it stays
between $6.7\times10^{-5}$ and $1.4\times10^{-4}$, below the NR error estimated
from the two highest resolutions. At $q\simeq10$ it rises to
$3$--$4\times10^{-4}$. Figure~\ref{fig:dephasing_sxs} shows the phasing at
$q=20$: the dephasing at merger is $0.07$~rad.

\begin{table}[t]
\caption{\label{tab:sxs_highq} Maximum $(2,2)$ unfaithfulness of \LMR{} against
the $q \gtrsim 10$ nonspinning SXS simulations over $M/M_\odot\in[10,300]$ for
Advanced LIGO (aLIGO, zero-detuned high power, $f_{\rm low}=20$\,Hz) and
$M/M_\odot\in[10,500]$ for the Einstein Telescope (ET2026a~\cite{Dupletsa:2026jzs},
$f_{\rm low}=3$\,Hz). The last column is the NR error, the unfaithfulness
between the two highest resolutions.}
\begin{ruledtabular}
\begin{tabular}{lcccc}
SXS & $q$ & $\bar{\mathcal{F}}^{\rm aLIGO}_{\max}$ & $\bar{\mathcal{F}}^{\rm ET}_{\max}$ & $\bar{\mathcal{F}}^{\rm NR}_{\max}$ (aLIGO / ET) \\
\hline
0196 & 9.66  & $3.0\times10^{-4}$ & $3.9\times10^{-4}$ & $3\times10^{-6}$ / $3\times10^{-6}$ \\
0185 & 9.99  & $2.9\times10^{-4}$ & $3.8\times10^{-4}$ & $5\times10^{-6}$ / $5\times10^{-6}$ \\
1107 & 10.0  & $2.7\times10^{-4}$ & $2.9\times10^{-4}$ & $2\times10^{-5}$ / $2\times10^{-5}$ \\
2480 & 14.0  & $1.3\times10^{-4}$ & $6.7\times10^{-5}$ & $2.3\times10^{-4}$ / $2.4\times10^{-4}$ \\
2477 & 15.0  & $1.4\times10^{-4}$ & $8.3\times10^{-5}$ & $5.0\times10^{-3}$ / $4.9\times10^{-3}$ \\
2516 & 20.0  & $8.4\times10^{-5}$ & $8.0\times10^{-5}$ & $4.5\times10^{-4}$ / $4.5\times10^{-4}$ \\
\end{tabular}
\end{ruledtabular}
\end{table}

\begin{figure}[t]
\includegraphics[width=\columnwidth]{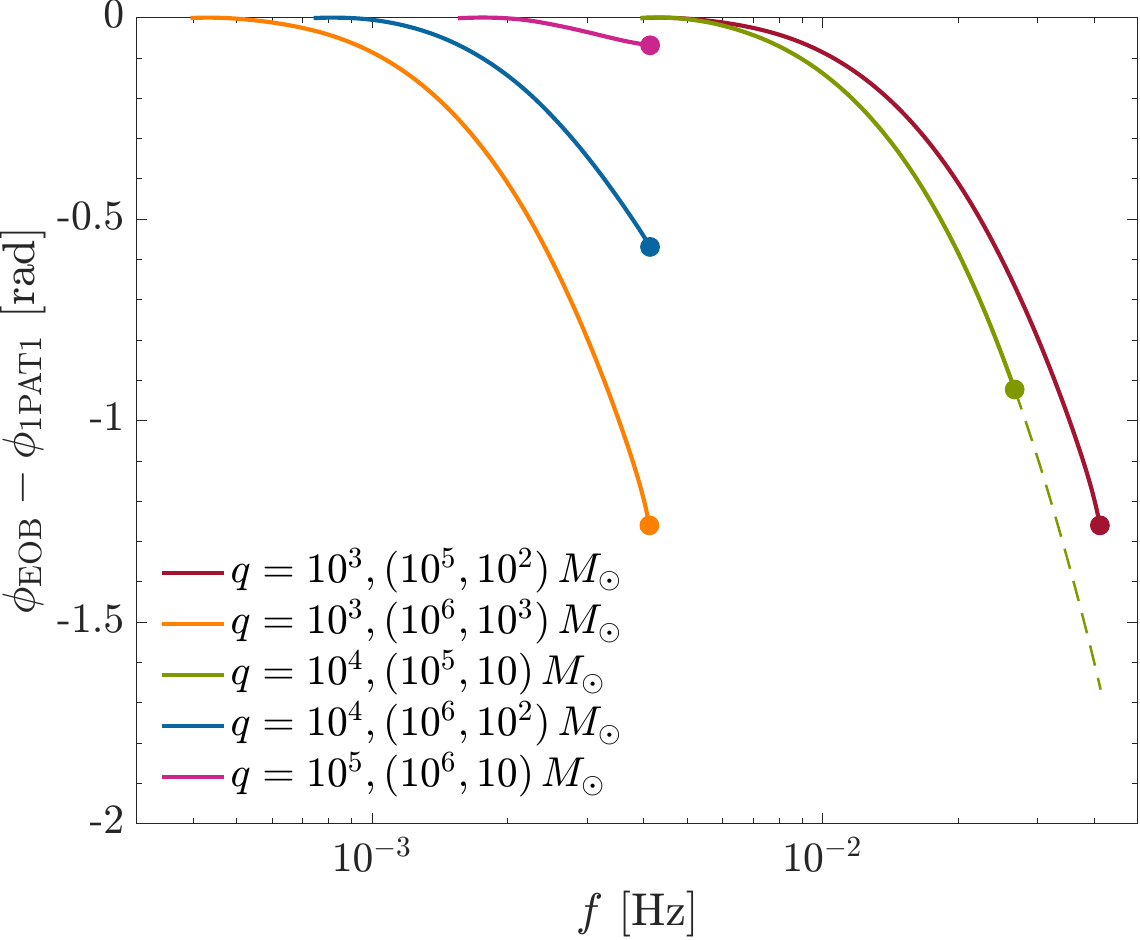}
\caption{\label{fig:1pat1_dphi} Accumulated $(2,2)$ dephasing between \LMR{}
and 1PAT1 versus frequency, for the configurations of Table~\ref{tab:1pat1_lisa}.
Curves are solid while the characteristic strain
($z=1$) is above the LISA noise and dashed after; the dot marks the in-band end.}
\end{figure}

\begin{figure}[t]
\includegraphics[width=\columnwidth]{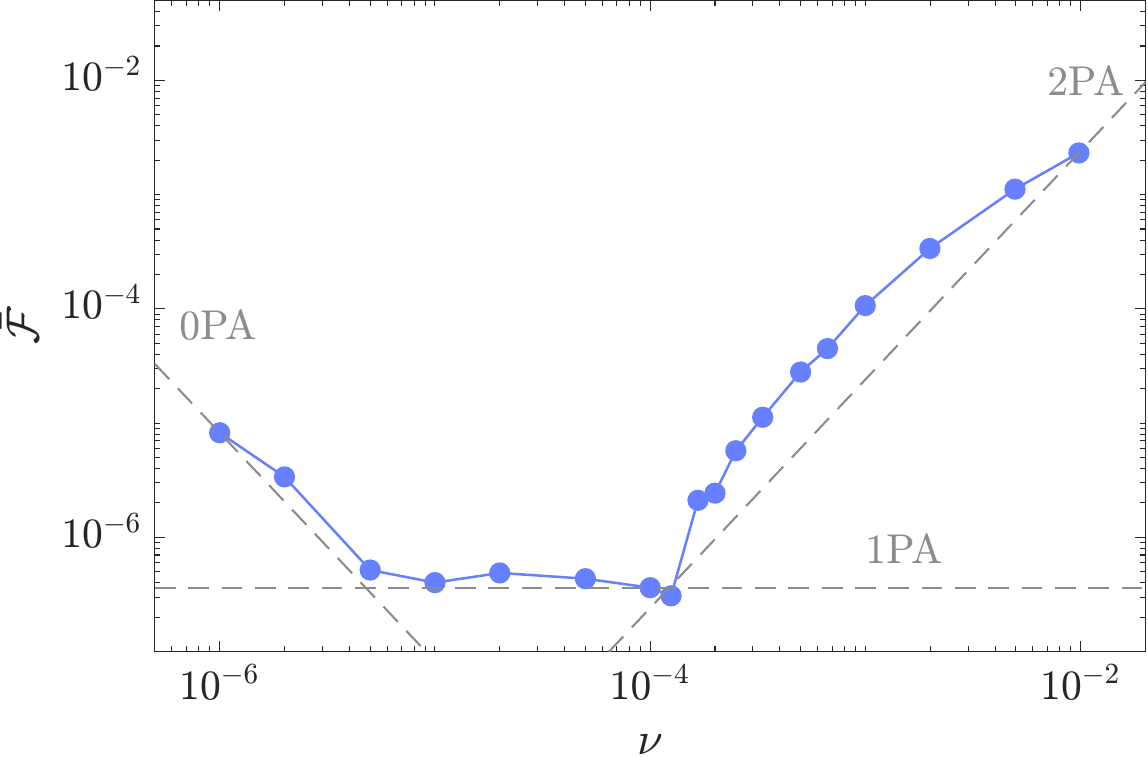}
\caption{\label{fig:sf_mismatch_nu} Flat-noise $(2,2)$ unfaithfulness
between \LMR{} and 1PAT1 versus $\nu$, for $q=10^2$--$10^6$.
The band runs from $M\omega_{22}=2\cdot10^{-3/2}$ ($r=10$) to the
end of 1PAT1. Dashed grey lines: 0PA ($\nu^{-2}$), 1PA ($\nu^{0}$) and 2PA
($\nu^{2}$) scalings, normalized at $q=10^6$, $10^4$ and $10^2$.}
\end{figure}

At larger mass ratios we compare with 1PAT1~\cite{Wardell:2021fyy}, generated with \texttt{WaSABI}~\cite{wardell_2026_18268703}, on the five LISA sources of
Table~\ref{tab:1pat1_lisa}, with $q=10^3$--$10^5$ and inspirals of up to four
years ending at the 1PAT1 stop. Figure~\ref{fig:1pat1_dphi} shows the accumulated dephasing versus frequency, and
Table~\ref{tab:1pat1_lisa} the unfaithfulness with the LISA sensitivity. Only two configurations exceed $1$~rad. At $q=10^3$, for both masses, the dephasing reaches $1.3$~rad, but 1PAT1 misses second-order post-adiabatic terms that matter there (Fig.~\ref{fig:sf_mismatch_nu}), so neither model is the reference; the $(10^5,10)\,M_\odot$ binary reaches $1.7$~rad only after leaving the LISA band.
Figure~\ref{fig:sf_mismatch_nu} shows the unfaithfulness against 1PAT1
for $q=10^2$--$10^6$ over a band fixed in radius, from $r=10$ to the end of
1PAT1, which
holds $\propto1/\nu$ cycles. An $n$PA error then dephases as $\nu^{n-1}$ and enters the unfaithfulness as
$\nu^{2(n-1)}$~\cite{Mitman:2025tmj}. The unfaithfulness follows the 2PA scaling up to
$q\simeq6\times10^3$, stays on a 1PA floor of $\sim4\times10^{-7}$ up to
$q\simeq2\times10^5$, set by the missing 5PN $O(\nu)$ flux, and rises as a 0PA error beyond. That residual stays below
$10^{-5}$ up to $q=10^6$, small enough for EMRIs: the four-year inspiral at
$q=10^5$ dephases by $0.07$~rad (Fig.~\ref{fig:1pat1_dphi}).

\begin{table}[t]
\caption{\LMR{} vs 1PAT1 for LISA configurations. $T$ is the duration of the compared 1PAT1 window,
which ends at the 1PAT1 stop ($r\simeq6.25$) and starts at $r_{\rm start}$ (1PAT1 requires $r_0 \le30$).
$\bar{\mathcal{F}}$ is the $(2,2)$ unfaithfulness
over the window band with the LISA Data Challenge SciRDv1 sky-averaged sensitivity including 4\,yr of galactic confusion noise. In parentheses: $\bar{\mathcal{F}}$ up to the in-band end ($26.6$~mHz).}
\label{tab:1pat1_lisa}
\begin{ruledtabular}
\begin{tabular}{ccccc}
$q$ & $(m_1, m_2)/M_\odot$ & $T$ [yr] & $r_{\rm start}$ & $\bar{\mathcal{F}}$ \\
\hline
$10^3$ & $(10^5, 10^2)$ & 0.23 & 30 & $3.1\times10^{-3}$ \\
$10^3$ & $(10^6, 10^3)$ & 2.35 & 30 & $3.3\times10^{-4}$ \\
$10^4$ & $(10^5, 10)$   & 2.37 & 30 & $2.0\times10^{-2}$ ($1.7\times10^{-2}$) \\
$10^4$ & $(10^6, 10^2)$ & 4.00 & 19.7 & $1.2\times10^{-4}$ \\
$10^5$ & $(10^6, 10)$   & 4.00 & 12.0 & $4.2\times10^{-5}$ \\
\end{tabular}
\end{ruledtabular}
\end{table}

\paragraph{Conclusions.}
We presented an effective-one-body model whose dynamics is not calibrated, covering $q>10$ up to the extreme-mass-ratio regime, through the intermediate range where neither NR nor self-force theory alone suffices.
Only the next-to-quasicircular corrections and the merger--ringdown, inherited from \Dali{}, are NR-informed.
Yet the model reaches unfaithfulness of $10^{-4}$--$10^{-5}$ against all SXS simulations with $q>10$, with
$0.07$~rad of dephasing at merger at $q=20$, and subradian dephasing against 1PAT1 over four-year LISA inspirals at
$q=10^4$ and $10^5$.

The key is the resummation of the main potential $A$: a Pad\'e
of the 6PN series plus the 5.5PN and 6.5PN contributions, with every known $\nu$-dependence, including the 5PN $\nu^2$
coefficient~\cite{Porto:2026fsd}, reproduces the first-order self-force $A$
function closely enough that the self-force data need not enter the model. The
analytical information now available is sufficient, once resummed this way, to
model large-mass-ratio binaries without calibration.

The model is also fast. A post-adiabatic inspiral carried down to the
transition to plunge and the fast frequency-domain path generate a full
$(2,2)$ waveform at $\sim60$~ms plus $\sim0.1\,\mu$s per frequency bin at any mass ratio, up to $\sim10^2\times$ faster
than an ODE evolution. IMRBs in next-generation detectors can thus be analyzed with an
inspiral--merger--ringdown model whose accuracy rests on analytical theory.

In passing, we note that the failure of the semi-analytical transition to plunge (see Speed) underlines the need for self-consistent time evolutions through the transition at intermediate mass ratios.

Everything validated here is quasicircular and nonspinning, with $q>10$, and
uses the $(2,2)$ mode; to build $h_+$ and $h_\times$ we recommend the $(2,2)$,
$(2,1)$, $(3,3)$ and $(4,4)$ modes. Comparable masses are covered by the NR-calibrated \Dali{}~\cite{Nagar:2024oyk, Albanesi:2025txj}. The model is also available for spinning binaries, with an uncalibrated spin sector (Hamiltonian and flux), currently reliable for primary spins $\lesssim0.6$, to be discussed in a follow-up study. Eccentric and precessing configurations, inherited from \Dali{}, are available but untested.

\paragraph{Acknowledgments.}
A.A. thanks Alessandro Nagar, Rossella Gamba, Simone Albanesi, Piero Rettegno and Danilo Chiaramello for useful discussions. A.A. also acknowledges IHES for hospitality during the development of this work.
A.A. and V.W. are supported by the Czech Science Foundation grant 26-23696S.

\TEOBResumS{} is developed open source and publicly available at

{\footnotesize \url{https://github.com/teobresums/teobresums}}.

The code version used here is tagged with the arXiv number associated to this work. \LMR{} is selected with the binary type \texttt{binary="LMR"} (model \Dali{}); it requires $q>10$.

The code is interfaced to state-of-art gravitational-wave data-analysis pipelines:
\href{https://github.com/matteobreschi/bajes}{bajes}~\cite{Breschi:2021wzr}, \href{https://git.ligo.org/lscsoft/bilby}{bilby}~\cite{Ashton:2018jfp} and \href{https://pycbc.org/}{pycbc}~\cite{Biwer:2018osg}.

\section{End Matter}

\begin{figure}[t]
\includegraphics[width=\columnwidth]{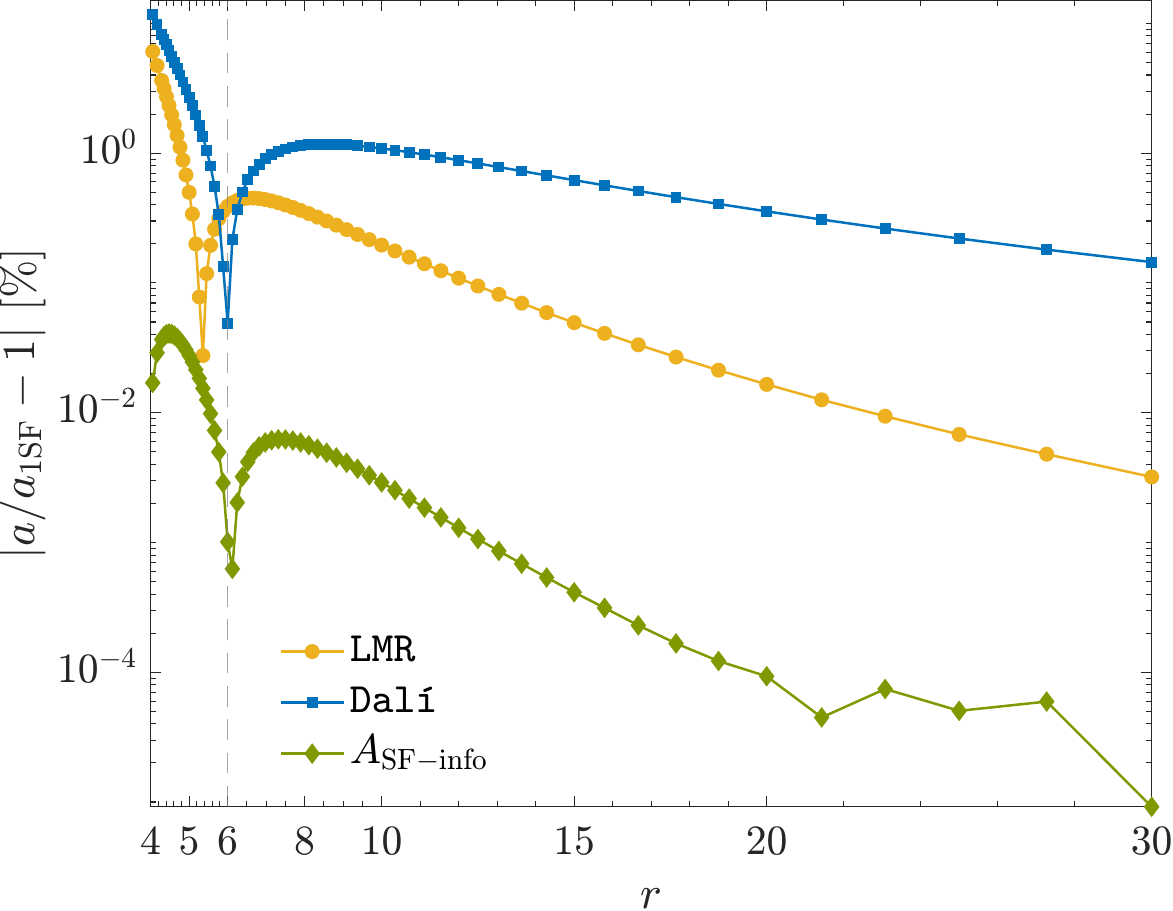}
\caption{\label{fig:a1sf_reldiff} Relative difference between the
linear-in-$\nu$ part $a(u)$ of $A$ and the exact 1SF function, versus $r=1/u$,
for \LMR{}, \Dali{} and the self-force-fitted $A$ of
Ref.~\cite{Nagar:2022fep} ($A_{\rm SF\text{-}info}$). The dashed line marks
$r=6$.}
\end{figure}

\paragraph{Comparing $A$ potential versions.}
\label{app:Acomp}
Figure~\ref{fig:a1sf_reldiff} compares the linear-in-$\nu$ part of
three $A$ potentials with the exact 1SF function, obtained from the 1SF
redshift~\cite{Akcay:2012ea} computed with
\texttt{WaSABI}~\cite{wardell_2026_18268703}. The $A$ of \LMR{}, Eq.~\eqref{eq:A_LMR},
built from analytical information only, stays within $0.45\%$ for $r\ge6.5$ and
$0.5\%$ down to $r=5$. The \Dali{} resummation of $A$~\cite{Nagar:2024oyk},
with a 5PN effective parameter calibrated on NR, deviates by up to $1.2\%$ at $r\simeq10$. The
$A$ of the previous large-mass-ratio version~\cite{Albertini:2024rrs}, informed with
numerical self-force data~\cite{Nagar:2022fep} (thus labeled $A_{\rm SF-info}$), matches it to $10^{-4}$, but is singular at the light
ring in the DJS gauge.

\begin{figure}[t]
\includegraphics[width=\columnwidth]{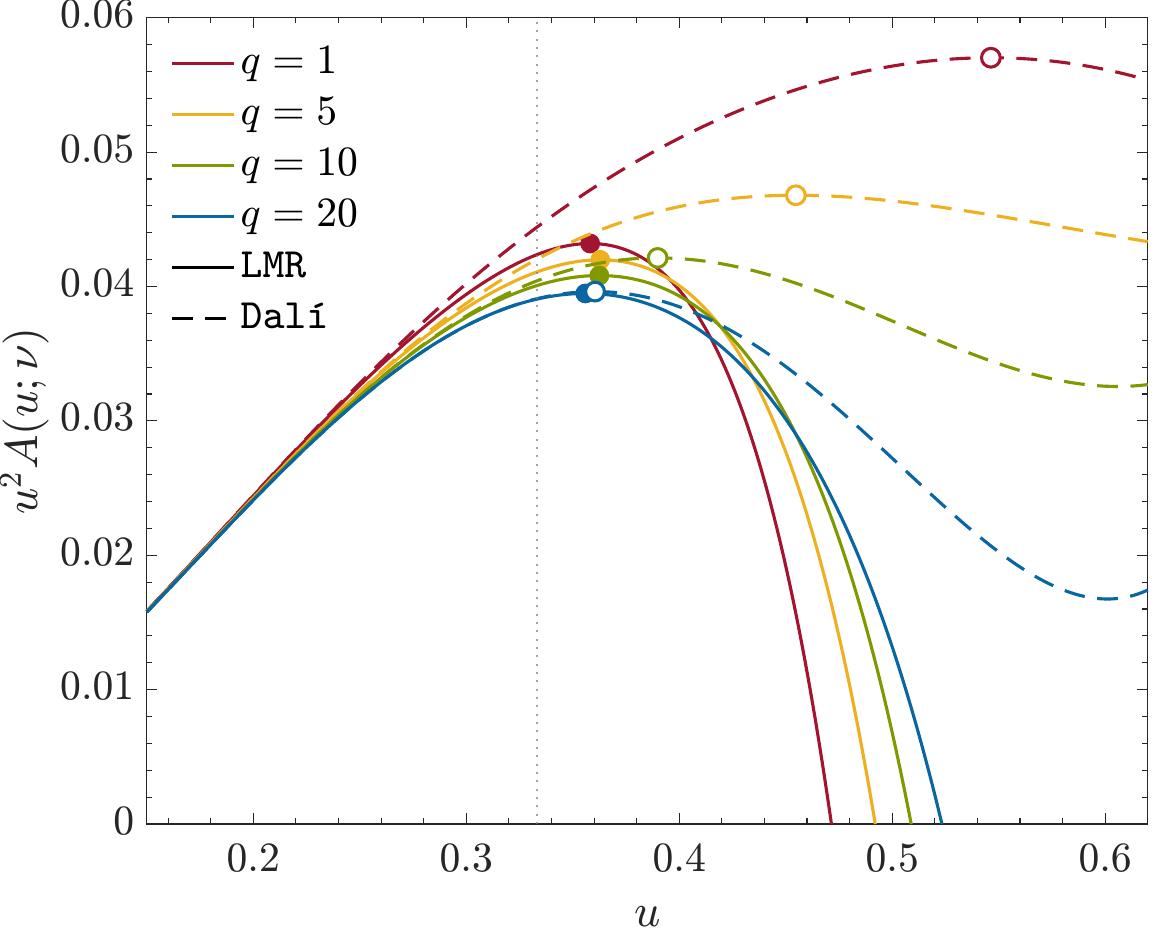}
\caption{\label{fig:u2A_LR} $u^2A(u;\nu)$ for \LMR{} (solid) and \Dali{}
(dashed) at $q=1$, $5$, $10$ and $20$. Circles mark the light ring, the maximum
of $u^2A$ (filled: \LMR{}; open: \Dali{}). The dotted line is the test-mass light
ring, $u=1/3$.}
\end{figure}

Figure~\ref{fig:u2A_LR} shows where each $A$ places the light ring, the
maximum of $u^2A$, for $q = {1,5,10,20}$. For \LMR{} it stays at $u\simeq0.36$ for every $q$. For
\Dali{} it moves from $u=0.36$ at $q=20$ to $u=0.55$ at $q=1$. The two agree at
$q=20$ and differ by $0.03$ at $q=10$. The half-integer terms of
Eq.~\eqref{eq:A_LMR}, known only at linear order in $\nu$, hold the \LMR{} light
ring in at comparable masses. Without them the light-ring structure is better
there, but the unfaithfulness against NR stays flat with mass ratio; with them
it decreases with $q$ from $q=1$ to $20$, so we keep them and restrict the model
to $q>10$.

\paragraph{Details on the $\rho_{22}$ at 4PN.}
\label{app:rho22}
Factorizing the 4PN $(2,2)$ mode of Ref.~\cite{Blanchet:2023sbv} [their
Eq.~(6.17)] as in Ref.~\cite{Damour:2008gu} gives the 4PN coefficient of
$\rho_{22}$,
\begin{equation}
\label{eq:rho22_4PN}
\rho_{22}^{(4)}\big|_{x_{22}} = \left(\frac{9202}{2205}+\frac{8819}{441}\,\nu\right)
\mathrm{eulerlog}_2(x) + \sum_{k=0}^{4} c_k\,\nu^k ,
\end{equation}
with $\mathrm{eulerlog}_2(x)=\gamma_E+2\ln 2+\tfrac12\ln x$ and
\begin{align}
c_0 &= -\frac{387216563023}{160190110080},\\
c_1 &= -\frac{6718432743163}{145627372800}-\frac{9953\,\pi^2}{21504},\\
c_2 &= \frac{10815863492353}{640760440320}-\frac{3485\,\pi^2}{5376},\\
c_3 &= -\frac{2088847783}{11650189824},\qquad
c_4 = \frac{70134663541}{512608352256}.
\end{align}
No coefficient of $A$ beyond 3PN enters, so Eq.~\eqref{eq:rho22_4PN} does not
depend on the 4PN and 5PN $A$ coefficients. %

The PN variable of Ref.~\cite{Blanchet:2023sbv} is built from the half-frequency
of the $(2,2)$ mode, $x_{22}$, which at 4PN differs from the orbital $x$ through
the tail-induced phase modulation [their Eq.~(6.10)],
\begin{align}
x_{22} &= x\left[1+\nu x^4 D(x)\right],\\
D &= -\frac{32}{15}-\frac{128}{5}\gamma_E-\frac{256}{5}\ln 2
     -\frac{128}{5}\ln\frac{b_0}{M}-\frac{192}{5}\ln x ,
\end{align}
with $b_0$ an arbitrary length scale. The phase of the factorized EOB mode,
$\arg T_{22}+\delta_{22}$, reproduces $D$ exactly with $b_0=2M/\sqrt{e}$, the
scale of the tail factor.

The EOB Hamiltonian reproduces the 4PN conservative energy $E_{\rm cons}(x)$,
whereas the flux-balance law reads
$\dot E_{\rm cons}+\dot E_{\rm Schott}=-\mathcal F_\infty$, with
$E_{\rm Schott}=-\tfrac12 M\nu^2x^5D$ at 4PN~\cite{Trestini:2025nzr}. The
radiation reaction must therefore carry $\mathcal F_\infty+\dot E_{\rm Schott}$
at the orbital $x$, where all $D$-dependent terms cancel:
\begin{equation}
\mathcal F_\infty+\dot E_{\rm Schott}
 = \mathcal F_\infty^{(x_{22})}(x)+\frac{6144}{25}\,\nu^3x^9 ,
\end{equation}
with $\mathcal F_\infty^{(x_{22})}$ the flux of Ref.~\cite{Blanchet:2023sbv}
[their Eq.~(6.11)] evaluated at $x_{22}=x$. Since $\hat f\propto\rho_{22}^4$,
the model uses
\begin{equation}
\label{eq:rho22_flux}
\rho_{22}^{(4)} = \rho_{22}^{(4)}\big|_{x_{22}}+\frac{48}{5}\,\nu ,
\end{equation}
which shifts $c_1$ only. With Eq.~\eqref{eq:rho22_flux} and the $\nu$-completed
$\rho_{\ell m}$ of the other modes, the re-expanded flux matches
$\mathcal F_\infty+\dot E_{\rm Schott}$ through 4.5PN.

The waveform uses the same $\rho_{22}$. Matching $|h_{22}|$ as a function of $x$
would instead require $\rho_{22}^{(4)}|_{x_{22}}+\tfrac12\nu D$, so
Eq.~\eqref{eq:rho22_flux} leaves a relative amplitude error
$2\nu x^4(48/5-D/2)$: $-3.0\times10^{-4}$ at $x=0.1$ and $+1.5\times10^{-3}$ at
$x=1/6$ for $\nu=1/4$, vanishing linearly as $\nu\to0$. A single $\rho_{22}$,
fixed by the flux, keeps the 4PN phasing exact.

\paragraph{Fold-aware switch to the ODEs.}
\label{app:fold}
We treat the right-hand sides of Eqs.~\eqref{eq:eom} as a black box on phase
space, so the construction survives future changes of the model. The LSO is the
``fold'' of the dynamical system where the stable and unstable circular branches merge, which is also the feature upon which the transition to plunge is centered~\cite{Ori:2000zn,Buonanno:2000ef,Compere:2021zfj,Kuchler:2024esj,RibesMetidieri:2026tat}:
$\dot p_{r_*}=\partial_r\dot p_{r_*}=0$ at $p_{r_*}=0$, equivalent to
$\partial_r\hat H_{\rm EOB}=\partial_r^2\hat H_{\rm EOB}=0$. Expanding about it
with $\delta r=r-r_{\rm LSO}\sim\nu^{2/5}$, $t-t_{\rm LSO}\sim\nu^{-1/5}$,
$p_{r_*}\sim\nu^{3/5}$, and the flux frozen at its LSO value, every neglected
term is suppressed by at least $\nu^{1/5}$, and the dynamics reduces to
$\ddot{\delta r}=-\mathcal A_{\rm P}\,\delta r^2-\mathcal B_{\rm P}\,(t-t_{\rm LSO})$,
with
$\mathcal A_{\rm P}=-\tfrac12(\partial_{p_{r_*}}\dot r)(\partial_r^2\dot p_{r_*})>0$
and
$\mathcal B_{\rm P}=-(\partial_{p_{r_*}}\dot r)(\partial_{p_\varphi}\dot p_{r_*})\hat{\mathcal F}_\varphi>0$
at the LSO. Rescaling $\delta r=-\Delta r_{\rm P}\,\zeta$,
$t-t_{\rm LSO}=t_{\rm P}T$ leaves the first Painlev\'e equation
$\zeta''=6\zeta^2+T$, with
$\Delta r_{\rm P}=6^{3/5}\mathcal A_{\rm P}^{-3/5}\mathcal B_{\rm P}^{2/5}$,
i.e.\ Eq.~\eqref{eq:DeltarP}, and
$t_{\rm P}=(6/\mathcal A_{\rm P}\mathcal B_{\rm P})^{1/5}\propto\nu^{-1/5}$. Far
before the LSO the solution follows the adiabatic branch
$\zeta=-\sqrt{-T/6}$; evaluating it at $T_{\rm sw}=-8$ gives
$c_{\rm sw}=\sqrt{-T_{\rm sw}/6}=\sqrt{4/3}$. The first neglected correction,
from the frozen flux, is
$\epsilon_{\rm P}=|\hat{\mathcal F}_\varphi|\,t_{\rm P}/p_{\varphi,{\rm LSO}}\sim\nu^{4/5}$;
the code refuses configurations with $\epsilon_{\rm P}>0.1$.

\begin{figure}[t]
\includegraphics[width=\columnwidth]{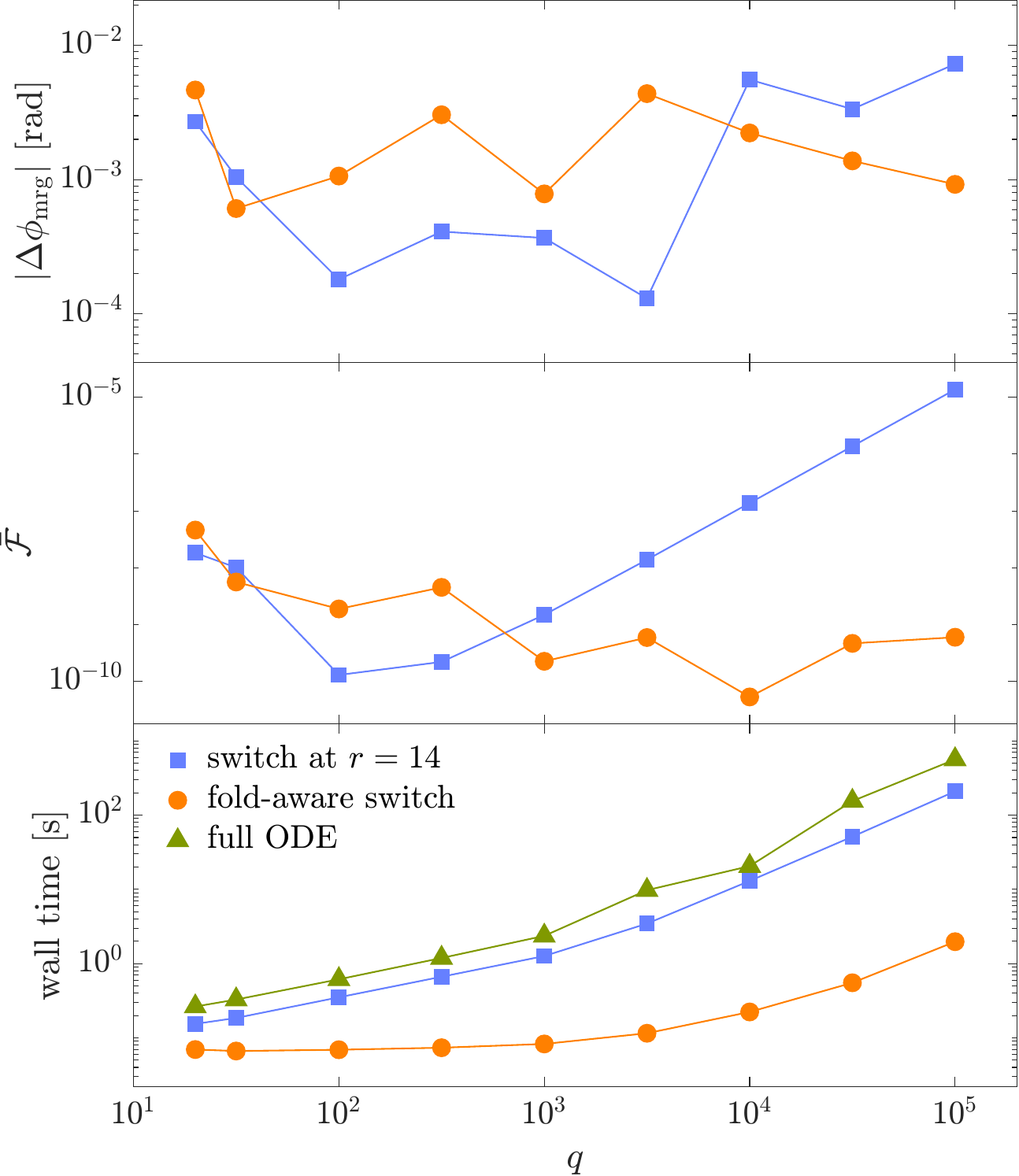}
\caption{\label{fig:handover} Post-adiabatic inspiral switched to the ODEs at
$r=14$ (squares) and at the fold-aware $r_{\rm switch}$ (circles), against
the full ODE evolution, all from $r_0=18$, versus $q$.
\textit{Top:} $(2,2)$ phase difference at merger, aligned from $r=16$;
\textit{middle:} flat-noise unfaithfulness over the whole waveform;
\textit{bottom:} single-core wall time, with the full ODE evolution (triangles).}
\end{figure}

The derivatives are fourth-order central finite differences of the implemented
right-hand sides (step $2\times10^{-3}$), and the fold conditions are solved by
a damped two-dimensional Newton iteration seeded at the test-mass innermost stable circular orbit. It
converges to ${\sim}10^{-11}$ in $r_{\rm LSO}$ within $\mathcal O(100)$
right-hand-side evaluations (${\sim}0.3$~ms, once per configuration). Nothing is
expanded in $\nu$: the resummed model enters only through pointwise evaluation.
Figure~\ref{fig:handover} compares the fold-aware switch with the $r=14$ switch of the previous large-mass-ratio version~\cite{Albertini:2024rrs}, against the full ODE evolution.

\bibliography{refs20261124}

\clearpage\onecolumngrid
\includepdf[pages=1]{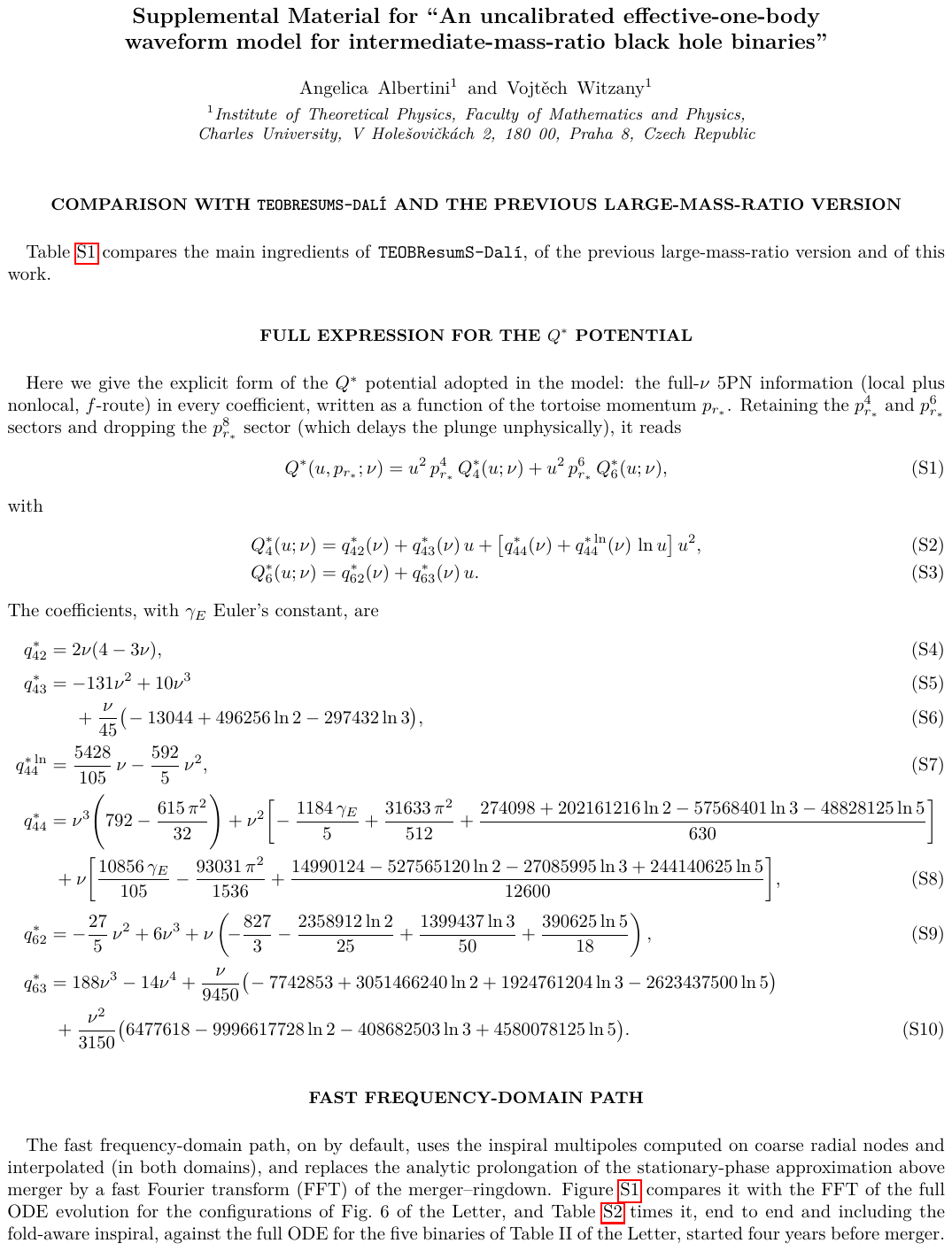}
\includepdf[pages=2]{supplemental.pdf}
\includepdf[pages=3]{supplemental.pdf}

\end{document}